\documentclass[a4paper]{article}
\begin{document}

\title{Regularized zero-range model and an application to the triton
and the hypertriton}

\author{D.V.Fedorov and A.S.Jensen \\
Aarhus University, 8000 Aarhus C, Denmark}

\date{}
\maketitle

\begin{abstract}
We examine the regularized zero-range model in an application to
three-fermion systems -- the triton and the hypertriton. We consider
bound states and low-energy neutron-deuteron and lambda-deuteron
scattering.  The model is shown to provide an adequate quantitative
description of these system on a par with finite-range potential models.
The well known correlation between the doublet $nd$ scattering length
and the triton binding energy (Phillips line) finds a natural explanation
within the model.

\end{abstract}

\section{Introduction}

The zero-range potential\footnote{also referred to as ``pseudo-potential''
or ``$\delta $-interaction''} is a useful concept which employs separation
of scales in physical systems \cite{demkov}.  The model allows selection
of the relevant degrees of freedom and provides qualitative and often
quantitative descriptions of intricate physical phenomena in an intuitive
and transparent way.  Conceptually it is equivalent to a coordinate
space formulation of effective field theories \cite{kolck}.

Particularly, the zero-range model effectively reduces a three-body
problem down to a one-body problem and thus provides a useful means
of model independent insights into the important properties of various
systems \cite{brown,nielsenmacek,fedichev,kartavtsev}.

However, a direct application of zero-range potentials to three-body
systems presents a problem -- a collapse of the system known as the
Thomas effect \cite{thomas}. To have a predictive power the zero-range
model needs to be regularized.

Several momentum-space regularization approaches have been recently
reported, which basically amount to a cut-off at high momentum
\cite{adhikari,amorim}, or a combination of a cut-off and an additional
three-body force \cite{bedaque}.

A combination of an energy cut-off and an effective three-body force
had also been used in a coordinate-space problem with
$\delta$-interactions \cite{esbensen}.

Recently we have introduced an alternative coordinate-space regularization
approach \cite{fed1jpa}. Unlike \cite{esbensen} we do not explicitly
use $\delta$-functions. The zero-range potential is formulated as a
boundary condition for the wave-function at the origin. Regularization
is achieved by a suitable generalization of the boundary condition and
can be thought of as a soft cut-off.

In contrast to momentum-space theories \cite{adhikari,amorim,bedaque}
our approach does not require an additional three-body parameter to set
the scale of the three-body binding energy.

The purpose of this paper is to test the suggested regularized zero-range
model in an application to systems of fermions, considering both bound
state problems and scattering processes, and comparing, where possible,
with finite-range potential models and effective field theories.

We shall give a brief introduction to the regularized zero-range
potential model, consider the hyper-spheric formalism of ''1+2''
scattering, and finally apply the model to the $nnp$ and $\Lambda np$
systems.

In the following we shall refer to the regularized zero-range potential
as contact interaction.

\section{Three-body problem with contact interactions}

In this chapter we discuss hyper-spheric adiabatic solution of the
three-body problem with contact interactions. We use the definition of
the two-body Schr\"{o}dinger equation with contact interaction as a
free $s$-wave equation with the solution $\psi_k(r)=\sin(kr+\delta)/r$,
and the boundary condition at the origin \cite{kolck,fed1jpa}
\begin{equation}
\left.\frac{1}{r\psi _{k}}\frac{d}{dr}r\psi_k\right|_{r=0}
=k\cot \delta =\frac{1}{a}+\frac{1}{2}Rk^{2}+PRk^{4} \;,  \label{bc2}
\end{equation}
where $\delta$ is the two-body phase shift, $a$, $R$, and $P$ are the
scattering length\footnote{note the sign convention for the scattering
length in (\ref{bc2})}, the effective range and the shape parameter and
$k$ is the wave-number. The $k^{2}$ and $k^{4}$ terms are necessary for
the contact interaction model to have a regular three-body ground state
solution \cite{fed1jpa}.

\subsection{Hyper-spheric adiabatic wave-function}

We use the \emph{hyper-spheric adiabatic approximation} \cite{review}
which employs the hyper-spheric coordinates $\rho$ and $\Omega$ (see
the appendix). The total wave-function $\Psi$ of a three-body system
is written as a product of a hyper-angular function $\Phi(\rho,\Omega)$
and a hyper-radial function $f(\rho)$
\begin{equation}
\Psi =\rho ^{-5/2}f(\rho )\Phi(\rho ,\Omega )\;.  \label{3bwf}
\end{equation}
The function $\Phi(\rho,\Omega)$ is an adiabatic eigenfunction of
the three-body Hamiltonian with a fixed hyper-radius $\rho $. The
corresponding eigenvalue $\lambda (\rho )$ serves as an effective
potential for the hyper-radial function $f(\rho )$
\begin{equation}
\left( -\frac{\partial ^{2}}{\partial \rho ^{2}}+\frac{\lambda (\rho )+15/4}{
\rho ^{2}}-Q(\rho )-\frac{2m}{\hbar ^{2}}E\right) f(\rho )=0,  \label{radial}
\end{equation}
where $E$ is the total energy of the three-body system, $m$ is the mass unit
used in the definition of $\rho $ (see the appendix), and 
\begin{equation}
Q(\rho )=\int d\Omega \Phi^{*}(\rho ,\Omega )\frac{\partial ^{2}
}{\partial \rho ^{2}}\Phi(\rho ,\Omega )\;.
\end{equation}

In most cases the one-channel adiabatic approximation (\ref{3bwf})
proves to be quite accurate \cite{esbenJPB}. However, for higher
accuracy several adiabatic eigenfunctions might need to be included.
Additional eigenfunctions are also needed, for example, for a description
of break-up processes. Although inclusion of additional eigenfunctions
is straightforward, in the present context we restrict ourselves to the
lowest adiabatic channel only.

\subsection{Adiabatic eigenvalue equation.}

We assume that contact interactions act only on the s-waves and
vanish everywhere except at the origin. The adiabatic eigenfunction
$\Phi(\rho,\Omega)$ can be then suitably expressed in terms of the
\emph{free} s-wave Faddeev components $\varphi _{\nu }(\alpha _{i})$
\begin{eqnarray}
\Phi(\rho ,\Omega ) &=&\sum_{i=1}^{3}A_{i}(\rho )\frac{\varphi
_{\nu }(\alpha _{i})}{\sin (2\alpha _{i})}\;, \\
\varphi _{\nu }(\alpha _{i}) &=&\sin \left[ \nu (\rho )\left( \alpha _{i}-
\frac{\pi }{2}\right) \right] \;,  \label{freefad}
\end{eqnarray}
where $\nu =\sqrt{\lambda +4}$, and $\alpha _{i}$ is the $i$-th
hyper-angle.  In the following with shall refer to both $\nu$ and
$\lambda=\nu^2-4$ as angular eigenvalues.

The quantities $A_{i}(\rho )$ and $\nu (\rho )$ are to be determined
from the boundary conditions (\ref{bc2}). The latter can be conveniently
reformulated in terms of the hyper-angles as \cite{fed1jpa}
\begin{eqnarray} \label{bc3}
&&\left[ \frac{1}{(2\alpha _{i}\Phi)}\frac{\partial }{
\partial \alpha _{i}}(2\alpha _{i}\Phi)\right] _{\alpha
_{i}=0} =  \nonumber \\
&&
\frac{\rho }{\sqrt{\mu _{i}}}\left[ \frac{1}{a_{i}}+\frac{1}{2}R_{i}\left( 
\frac{\sqrt{\mu _{i}}\nu }{\rho }\right) ^{2}+P_{i}R_{i}^{3}\left( \frac{
\sqrt{\mu _{i}}\nu }{\rho }\right) ^{4}\right] \;,
\end{eqnarray}
where $a_{i}$, $R_{i}$, $P_{i}$ and $\mu _{i}$ are the scattering length,
effective range, shape parameter and the reduced mass (divided by the mass
unit $m$) for the two-body system of particles $j$ and $k$. We assume here
and in the following that the indices $\{i,j,k\}$ form a cyclic permutation
of \{1,2,3\}.

For small $\alpha _{i}$ the wave-function $\Phi$ can be expanded
as \cite{review}
\begin{eqnarray}
\sin (2\alpha _{i})\Phi &=&A_{i}(\rho )\varphi _{\nu }(\alpha
_{i})\nonumber \\
&+&A_{j}(\rho )2\alpha _{i}\frac{\varphi _{\nu }(\phi _{ji})}{\sin (2\phi
_{ji})}+A_{k}(\rho )2\alpha _{i}\frac{\varphi _{\nu }(\phi _{ki})}{\sin
(2\phi _{ki})}
+O(\alpha _{i}^{2})\;, \\
\phi _{ji} &=&\arctan \left(
\sqrt{\frac{m_{k}(m_{1}+m_{2}+m_{3})}{m_{j}m_{i}} }\right) \;.
\end{eqnarray}

The boundary conditions (\ref{bc3}) then turn into a set of three linear
equations for the coefficients $A_{i}(\rho )$ 
\begin{eqnarray}\label{MA0}
&&\sum_{j=1}^{3}M_{ij}A_{j}(\rho )=0\;, \\
&&M_{ii}=\varphi _{\nu }^{\prime }(0)-\varphi _{\nu }(0)L_{i}\;,
 \nonumber \\
&&M_{ij}=2\frac{\varphi _{\nu }(\phi _{ji})}{\sin (2\phi _{ji})}\;,\;i\neq
j\;,  \nonumber
\end{eqnarray}
where $\varphi _{\nu }^{\prime }(0)\equiv ~\left. d\varphi _{\nu }(\alpha
)/d\alpha \right| _{\alpha =0}$ and where we have used a short-hand notation 
\begin{eqnarray}\label{L}
L_{i}\equiv \frac{\rho }{\sqrt{\mu _{i}}}\left[ \frac{1}{a_{i}}+\frac{1}{2}
R_{i}\left( \frac{\sqrt{\mu _{i}}\nu }{\rho }\right)
^{2}+P_{i}R_{i}^{3}\left( \frac{\sqrt{\mu _{i}}\nu }{\rho }\right)
^{4}\right] \;. 
\end{eqnarray}

The set of equations (\ref{MA0}) has nontrivial solutions only when
the determinant of the matrix $M(\rho ,\nu )$ is zero, which gives the
equation to determine $ \nu (\rho )$ and $\lambda (\rho )=\nu^2(\rho )-4$,
\begin{equation}
\det M(\rho ,\nu )=0\;.  \label{detm0}
\end{equation}

\section{''1+2'' scattering}

In this chapter we discuss the hyper-spheric formalism for the elastic
scattering of a particle against a bound system of the other two.
This process is often referred to as ''1+2'' scattering.

\subsection{Adiabatic solution at large distances}

A bound two-body subsystem -- say, number 3 -- with the binding energy $
B=\hbar^{2}\kappa _{3}^{2}/(2m\mu _{3})$ gives rise to an eigenvalue $
\nu(\rho)$ which at large $\rho$ asymptotically behaves as \cite{fed1jpa}
\begin{eqnarray}
\nu (\rho ) &=&i\frac{\rho }{\sqrt{\mu _{3}}}\kappa _{3},  \label{nyas} \\
\lambda (\rho ) &=&-\frac{2mB}{\hbar ^{2}}\rho ^{2}-4,
\end{eqnarray}
with the corresponding (normalized) adiabatic wave-function
(\ref{freefad}) being
\begin{equation}  \label{bawf}
\varphi _{\nu }(\alpha _{3})=\sqrt{2\frac{\rho \kappa _{3}}{\sqrt{\mu _{3}}}}
\exp \left( -\frac{\rho \kappa _{3}}{\sqrt{\mu _{3}}}\alpha _{3}\right) \;.
\label{e-a}
\end{equation}

Indeed, $\varphi _{\nu }(\phi _{2i})$ vanishes exponentially for large $\rho 
$ and the 3-rd equation in (\ref{bc3}) asymptotically decouples from the
other two and turns into 
\begin{equation}
i\nu =\frac{\rho }{\sqrt{\mu _{3}}}\left[ \frac{1}{a_{3}}+\frac{1}{2}
R_{3}\left( \frac{\sqrt{\mu _{3}}\nu }{\rho }\right)
^{2}+P_{3}R_{3}^{3}\left( \frac{\sqrt{\mu _{3}}\nu }{\rho }\right)
^{4}\right] \;,
\end{equation}
which with the substitution $\nu \rightarrow k\rho /\sqrt{\mu _{3}}$ takes
the form of the usual equation for the two-body bound state 
\begin{equation}
ik=k\cot \delta (k)
\end{equation}
with the solution $i\kappa _{3}$. Note that since $\varphi _{\nu }(\phi
_{3i})$ vanishes exponentially with $\rho $ the correction to (\ref{nyas})
is also exponentially small \cite{fed1jpa}.

The additional term $Q$ in the hyper-radial equation (\ref{radial})
asymptotically approaches $Q\rightarrow -1/(4\rho ^{2})$, which can be
directly verified from (\ref{e-a}). Although this term is generally small it
is important to ensure the correct asymptotics of the hyper-radial equation
in the case of a bound two-body subsystem \cite{fed1jpa}. For simplicity we
shall in the following always use only the leading term $-1/(4\rho ^{2})$,
often referred to as the Langer correction term. The hyper-radial
equation then becomes 
\begin{equation}
\left( -\frac{\partial ^{2}}{\partial \rho ^{2}}+\frac{\lambda (\rho )+4}{
\rho ^{2}}-\frac{2m}{\hbar ^{2}}E\right) f(\rho )=0,  \label{radial2}
\end{equation}
with the correct ''1+2'' asymptotics at large $\rho$
\begin{equation}
\left( -\frac{\partial ^{2}}{\partial \rho ^{2}}-\frac{2m}{\hbar ^{2}}
(E+B)\right) f(\rho )=0.  \label{ra}
\end{equation}

\subsection{Scattering wave-function and phase shift}

Within the contact interaction model the two-body bound state wave-function
simplifies to $\exp (-\kappa _{3}r_{12})/r_{12}$, where $r_{12}$ is the
relative coordinate. The wave-function describing an elastic scattering of
the third particle against the bound system of the first two must then
have the following asymptotic form
\begin{equation}\label{1p2}
\Psi \rightarrow \frac{1}{r_{12}}\exp (-\kappa _{3}r_{12})\frac{1}{r_{(12)3}}
\sin (kr_{(12)3}+\delta )\;,
\end{equation}
where $r_{(12)3}$ is the relative distance between the third particle and
the center of mass of the first two, $\delta $ is the sought ''1+2''
phase shift, and the scattering energy is $\hbar ^{2}k^{2}/(2m\mu _{3})$.

We shall show now that the three-body wave-function (\ref{3bwf}) indeed
takes this asymptotic form at large $\rho $. The asymptotic form (\ref{ra})
of the hyper-radial equation has a solution 
\begin{equation}
f(\rho )=\sin (q\rho +\Delta ) \;,  \label{hphase}
\end{equation}
where $\Delta (q)$ is the hyper-radial phase and
$q^2=2m(E+B)/\hbar^2>0$. Multiplying the hyper-radial function
(\ref{hphase}) and the adiabatic function (\ref{bawf}) gives the total
wave-function
\begin{equation}
\Psi =\rho ^{-5/2}\sin (q\rho +\Delta(q))\frac{1}{\sin (2\alpha_{3})}\sqrt{2 
\frac{\rho \kappa _{3}}{\sqrt{\mu _{3}}}}\exp \left( -\frac{\rho \kappa _{3} 
}{\sqrt{\mu _{3}}}\alpha _{3}\right) \;.  \label{wfra}
\end{equation}
For $\rho\kappa_{3}/\sqrt{\mu_{3}}\gg 1$ the exponent in (\ref{wfra}) is
non-vanishing only when $\alpha _{3}\ll 1$. Therefore in this region $\rho
\alpha _{3}=x$ and $\rho =y$ (neglecting terms of the order of $\alpha
_{3}^{2}$) and the wave-function (\ref{wfra}) becomes
\begin{equation}
\Psi \simeq \frac{1}{x}\exp \left( -\frac{\kappa _{3}}{\sqrt{\mu _{3}}}
x\right) \frac{1}{y}\sin (qy+\Delta(q)) \;.
\end{equation}
Recalling that $x=\sqrt{\mu_3}r_{12}$ and $y=\sqrt{\mu_{(12)3}}r_{(12)3}$
(see the appendix) we obtain
\begin{equation}
\Psi \simeq \frac{1}{r_{12}}\exp \left( -\kappa_{3}r_{12}\right) \frac{1}{
r_{(12)3}}\sin (kr_{(12)3}+\Delta(q)) \;,
\end{equation}
where $k=q\sqrt{\mu_{(12)3}}$. This is exactly the sought asymptotic form
(\ref{1p2}).  The hyper-radial phase shift $\Delta(q)$ is then equal
to the ''1+2'' phase shift $\delta(k)$.

Thus the ''1+2'' phase shift $\delta (k)$ can be easily calculated
within the hyper-spheric approach by solving the hyper-radial equation
(\ref{radial}) with the wave-vector $q=k/\sqrt{\mu_{(12)3}}$ and
extracting the hyper-radial phase $\Delta (q)$ from the asymptotic form
(\ref{hphase}). The ''1+2'' phase shift $\delta (k)$ is equal to the
hyper-radial phase $\Delta (q)$. The ''1+2'' scattering length can then
be found as the limit $a^{-1}=\lim_{k\rightarrow 0}k\cot\delta$.

\section{Contact interactions for $np$ and $nn$ systems}

The basic quantities that characterize the $np$ interaction at low energies
are the phase shifts and the deuteron binding energy. We therefore choose the
contact interaction parameters $a$, $R$, and $P$ in (\ref{bc2}) by fitting
the phase shifts while keeping the deuteron binding energy fixed at -2.2~MeV.
For the $nn$ system we instead keep the energy of the virtual $s$-state at
-0.1~MeV. Our parametrization does not allow change of sign of the phase
shifts at large energies. Therefore we can only aim at describing the phases
below 100~MeV for $nn$ and below 150~MeV for $np$ scattering.

\begin{table}
\caption{
Parameters of the contact interaction (\ref{bc2}) in the $^3S_1$ (triplet)
$np$ and $^{1}S_{0}$ (singlet) $nn$ channels. Here $a$ is the scattering
length, $R$ -- the effective range, $P$ -- the shape parameter, $E_{d}$
is the energy of the bound $np$ state (deuteron), and $E_{virt}$ is the
energy of the virtual (anti-bound) $nn$ state.  }

\begin{center}
\begin{tabular}{|c|c|c|c|c|}
\hline
$V_{np}(^{3}S_{1})$ & $a$, fm & $R$, fm & $P$ & $E_{d}$, MeV \\ 
\hline
I & -5.26 & 1.54 & 0.084 & -2.2 \\ 
II & -5.14 & 1.38 & 0.143 & -2.2 \\ 
III & -5.02 & 1.21 & 0.256 & -2.2 \\ 
\hline
$V_{nn}(^{1}S_{0})$ & $a$, fm & $R$, fm & $P$ & $E_{virt}$, MeV \\ 
\hline
I & 19.00 & 2.10 & 0.086 & -0.1 \\ 
II & 19.09 & 1.91 & 0.132 & -0.1 \\ 
III & 19.21 & 1.64 & 0.238 & -0.1 \\
\hline
\end{tabular}
\end{center}
\label{NN}\end{table}

For each pair of particles we consider three sets of parameters labelled I,
II, and III (see Table 1), where II is more or less the best possible fit,
while I and III represent small variations towards better description of
correspondingly lower and higher energies.

\begin{figure}
\begin{center}
\input{fignp.tex}
\end{center}
\caption{
The $^{3}S_{1}$ $np$ phase shift $\delta$ as function of
the center-of-mass energy $E_{cm}$ for the contact interactions from
Table 1.  The circles represent the results from the Argonne $v_{18}$
potential \cite{v18}.  }
\label{np_phase}\end{figure}

\begin{figure}
\begin{center}
\input{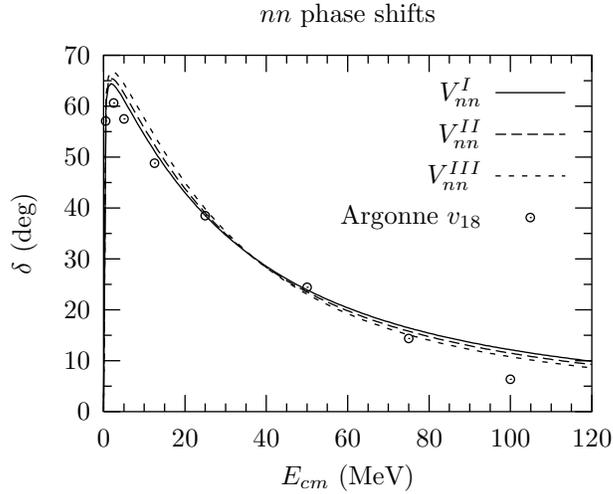}
\end{center}
\caption{
The $^{1}S_{0}$ $nn$ phase shift $\delta $ as function of the
center-of-mass energy $E_{cm}$ for the contact interactions from Table 1.
The circles represent the results from the Argonne $v_{18}$ potential
\cite{v18}.  }
\label{nn_phase}\end{figure}

The $np$ and $nn$ phase shifts for these models are shown,
correspondingly, on Fig.~\ref{np_phase} and Fig.~\ref{nn_phase}. We have
used $\hbar c$=197.3~MeV~fm and $m_{n}$=$m_{p}$ =939~MeV. Potentials
I provide a better description at lower energies while overestimate
the phases at larger energies. Potentials III are better in the higher
energy region at the expense of the lower energies, while potentials II
represent a reasonable compromise.

The larger $P$-parameters give smaller phase shifts at higher energy or,
correspondingly, weaker interaction. One can say that the $P$-parameter
smoothly cuts off the interaction at higher energies, thus regularizing
the three-body problem \cite{fed1jpa}.

\section{Application to the $nnp$ system}

The spins of three nucleons can couple to a total spin $S$ of 3/2 and
1/2, which for relative $s$-waves is also the total angular momentum. In
the $S$ =1/2 (doublet) case there is a bound state of a proton and two
neutrons -- a triton. No three-body bound state is found in the $S$=3/2
(quartet) case. We assume that the neutron and the proton have the same
mass $m$=939~MeV, which is also used as the mass scale in the definition
of $\rho$ (see appendix).

\subsection{Quartet $nd$ scattering}

\subsubsection{Adiabatic eigenfunction}

In the case of particles with spins we have to include also the
spin-functions $\zeta $ in the adiabatic wave-function 
\begin{equation}\label{Phspin}
\Phi (\rho ,\Omega )=\sum_{i=1}^{3}\zeta _{i}A_{i}\frac{\varphi _{\nu
}(\alpha _{i})}{\sin (2\alpha _{i})}\;.  \label{qphi}
\end{equation}
For $S$=3/2 the spin functions $\zeta _{i}$ are represented by 
\begin{equation}
\zeta _{i}=\uparrow _{1}\uparrow _{2}\uparrow _{3}\;,
\end{equation}
where $\uparrow _{i}$ designates the spin-function of the $i$-th particle
where the $z$-projection of the spin operator is equal to +1/2.

Let the proton be particle number 1 and the neutrons number 2 and 3. Two
neutrons with total spin 1 can not be in an s-state. Therefore the Faddeev
component number 1 must vanish identically. Under exchange of the neutrons
the angular functions $\varphi_{\nu}(\alpha_2)$ and $\varphi_{\nu}(\alpha_3)$
in (\ref{qphi}) transform into each other while the spin functions $
\zeta_{i} $ do not change. Therefore the antisymmetric wave-function for $S$
=3/2 is
\begin{equation}
\Phi (\rho ,\Omega )=A_{2}\zeta _{2}\left( \frac{\varphi _{\nu }(\alpha _{2})
}{\sin (2\alpha _{2})}-\frac{\varphi _{\nu }(\alpha _{3})}{\sin (2\alpha
_{3})}\right) \;.
\end{equation}

\subsubsection{Eigenvalue equation}

The system of three equations (\ref{MA0}) reduces in this case to only one
equation which includes triplet neutron-proton parameters
\begin{equation}
\varphi_\nu^\prime(0)-\frac{4}{\sqrt{3}}\varphi _{\nu }(\frac{\pi}{3})
=L_{np}\varphi _{\nu }(0) \;,  \label{q}
\end{equation}
where we have used that $\phi _{ji}=\pi /3$ for three particles with
the same mass and where $L_{np}$, see eq.~(\ref{L}), contains triplet
neutron-proton parameters $a_{np}$, $R_{np}$, and $P_{np}$
\begin{equation}
L_{np}=\frac{\rho }{\sqrt{\mu }}\left[ \frac{1}{a_{np}}+\frac{1}{2}
R_{np}\left( \frac{\sqrt{\mu }\nu }{\rho }\right)
^{2}+P_{np}R_{np}^{3}\left( \frac{\sqrt{\mu }\nu }{\rho }\right) ^{4}\right]
.\end{equation}
Using the explicit expressions (\ref{freefad}) for the functions $\varphi
_{\nu }$ the eigenvalue equation (\ref{q}) can be written as
\begin{eqnarray}  \label{qnu}
\frac{\nu \cos (\nu \frac{\pi }{2})+\frac{4}{\sqrt{3}}\sin (\nu \frac{\pi }{6
})}{-\sin (\nu \frac{\pi }{2})}\nonumber \\
=\frac{\rho }{\sqrt{\mu }}\left[ \frac{1}{
a_{np}}+\frac{1}{2}R_{np}\left( \frac{\sqrt{\mu }\nu }{\rho }\right)
^{2}+P_{np}R_{np}^{3}\left( \frac{\sqrt{\mu }\nu }{\rho }\right) ^{4}\right]
\;,  \label{qlam}
\end{eqnarray}
where we have dropped the index at the reduced mass $\mu $ as we assumed
that all three nucleons have the same mass.

For small $\rho $ this equation has a root which to lowest orders in
$\rho$ is 
\begin{eqnarray}
\nu &=&4-\frac{3}{64\left( \sqrt{\mu }\right) ^{3}}\frac{\rho ^{3}}{\pi
P_{np}R_{np}^{3}}\;, \\
\lambda &=&12-\frac{3}{8\left( \sqrt{\mu }\right) ^{3}}\frac{\rho ^{3}}{\pi
P_{np}R_{np}^{3}}\;.
\end{eqnarray}
The solution $\lambda $(0)=12 is an ordinary solution for regular
finite-range potentials \cite{review} corresponding to the hyper-angular
quantum number $K$=2.

When $R_{np}$ and $P_{np}$ are equal to zero, as in the ordinary zero-range
potential, the eigenvalue equation (\ref{qnu}) at $\rho \ll a_{np}$ has a 
\emph{real} root $\nu\cong$2.166. Therefore there are no Thomas or Efimov
effects in the quartet case even without regularization \cite{fed1jpa}.

For large distances, due to the bound two-body subsystem (deuteron)
the eigenvalue is asymptotically approaching
\begin{eqnarray}\label{as}
\nu &\rightarrow &i\frac{\rho }{\sqrt{\mu }}\kappa _{d}\ \;,  \label{br2} \\
\lambda &\rightarrow &-\frac{2mB_{d}\rho ^{2}}{\hbar ^{2}}\ -4
\;,
\end{eqnarray}
where $B_{d}$=2.2~MeV is the deuteron binding energy and $\kappa _{d}=\sqrt{
2m\mu B_{d}/\hbar ^{2}}$ ($\kappa_d\cong$0.23~fm$^{-1}$).

\subsubsection{Adiabatic eigenvalues and results}

\begin{figure}
\begin{center}
\input{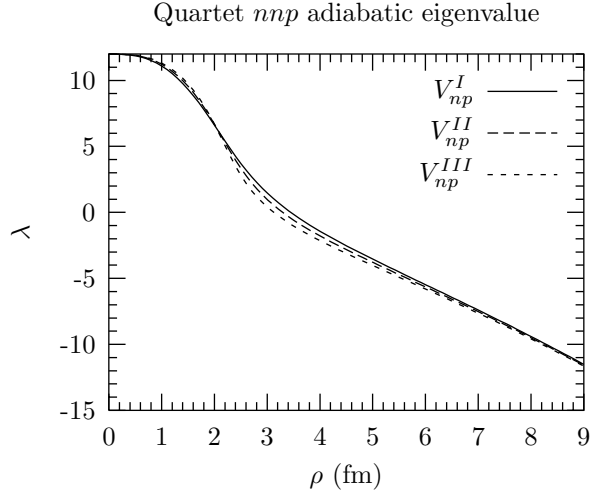}
\end{center}
\caption{
Quartet adiabatic eigenvalue $\lambda$ as a function of $\rho$
for the three $np$ interactions from Table 1.
}
\label{qlfig}
\end{figure}

The solutions $\lambda(\rho)$ of the equation (\ref{qlam}) for the three $
np$ potentials from Table 1 are shown on Fig.~\ref{qlfig}. The difference
between the curves is not big and, moreover, the region, where the curves
differ appreciably, is largely limited to $\rho\le$5~fm. However, in
this region $ \lambda +4>$0 and, consequently, the effective potential
$(\lambda +4)/\rho ^{2}$ in the hyper-radial equation (\ref{radial2})
is repulsive. Therefore the hyper-radial wave-function in this region
is small and thus insensitive to the small variations in the eigenvalues
$\lambda $.

The quartet $nd$ scattering lengths calculated using these three adiabatic
eigenvalues in the hyper-radial equation (\ref{radial2}) are given in
Table 2. All three contact interaction models give similar results
which are very close to the experiment and to the result of Bedaque
and van Kolck \cite{bedaquePLB98} obtained within the effective field
theory approach.

\begin{table}
\caption{
Quartet $nd$ scattering length, $^{4}a_{nd}$. Here $V_{np}$ refers
to the $np$ contact interaction from Table 1.
}
\begin{center}
\begin{tabular}{|c|ccc|c|c|}
\hline
& $V_{np}^{(I)}$ & $V_{np}^{(II)}$ & $V_{np}^{(III)}$ &EFT \cite{bedaquePLB98}& 
Exp. \cite{dilg} \\ 
\hline
$^{4}a_{nd}$, fm & -7.03 & -6.86 & -6.67 & -6.33 & -6.35$\pm $0.02 \\
\hline
\end{tabular}
\end{center}
\end{table}

\subsection{Doublet $nd$ scattering and the triton}

\subsubsection{Adiabatic eigenfunction}

For the doublet ($S$=1/2) state we have to include in the adiabatic
function (\ref{Phspin}) three different spin-functions, $\sigma_{nn}$,
$\tau_{_2p}$, $\tau_{p3}$ where the two neutrons, $n_{1}$ and $n_{2}$,
are in the singlet state while the neutron-proton pairs,  $n_{2}p$ and
$pn_{1}$ are in the triplet state
\begin{eqnarray}
\sigma _{nn} =\left( \left( n_{1}\otimes n_{2}\right) _{0}\otimes p\right)
_{\frac{1}{2}}=\frac{1}{\sqrt{2}}\left( \uparrow _{2}\downarrow
_{3}-\downarrow _{2}\uparrow _{3}\right) \uparrow _{1},  \nonumber
\label{ksit} \\
\tau_{_2p} =\left( \left( n_{2}\otimes p\right) _{1}\otimes
n_{1}\right) _{\frac{1}{2}}=\sqrt{\frac{2}{3}}\left( \uparrow _{3}\uparrow
_{1}\right) \downarrow _{2}-\sqrt{\frac{1}{6}}\left(
\uparrow _{3}\downarrow _{1}+\downarrow _{3}\uparrow _{1}\right) \uparrow
_{2}, \\
\tau _{p3} =\left( \left( p\otimes n_{1}\right) _{1}\otimes n_{2}\right) _{
\frac{1}{2}}=\sqrt{\frac{2}{3}}\left( \uparrow _{1}\uparrow _{2}\right)
\downarrow _{3}-\sqrt{\frac{1}{6}}\left( \uparrow
_{1}\downarrow _{2}+\downarrow _{1}\uparrow _{2}\right) \uparrow _{3}, 
\nonumber
\end{eqnarray}
where $(a\otimes b)_{s}$ denotes coupling of the spins of the particles (or
systems of particles) $a$ and $b$ to the total spin $s$.

The wave-function must be antisymmetric under exchange of the two identical
neutrons, which in our case are particles number 2 and 3. Under
transformation 2$\longleftrightarrow $3 the wave-functions undergo the
following transformations 
\begin{eqnarray}
&\varphi _{\nu }(\alpha _{1}) \rightarrow \varphi _{\nu }(\alpha
_{1}),\;\varphi _{\nu }(\alpha _{2})\longleftrightarrow \varphi _{\nu
}(\alpha _{3})& \\
&\sigma _{nn} \rightarrow -\sigma _{nn},\;\tau _{2p}\longleftrightarrow
\tau _{p3} \nonumber \;,&
\end{eqnarray}
where $\varphi _{\nu }(\alpha _{i})$ are the usual solutions (\ref{freefad})
to the free Faddeev equations. The antisymmetric hyper-angular function can
therefore be written as 
\begin{equation}\label{wft}
\Phi (\rho ,\Omega )=\sigma _{nn}A_{1}\frac{\varphi _{\nu }(\alpha _{1})}{
\sin 2\alpha _{1}}+\tau _{2p}A_{2}\frac{\varphi _{\nu }(\alpha _{2})}{\sin
2\alpha _{2}}-\tau _{p3}A_{2}\frac{\varphi _{\nu }(\alpha _{3})}{\sin
2\alpha _{3}}  \label{dphi}
\;. \end{equation}

\subsubsection{Eigenvalue equation}

The hyper-angular function must satisfy the spin-projected boundary
conditions 
\begin{eqnarray}
\frac{\partial }{\partial \alpha _{1}}\left[ \alpha _{1}\left( \sigma
_{nn}^{\dagger }\Phi \right) \right] _{\alpha _{1}=0} &=&L_{nn}\left[ \alpha
_{1}\left( \sigma _{nn}^{\dagger }\Phi \right) \right] _{\alpha _{1}=0}, \\
\frac{\partial }{\partial \alpha _{2}}\left[ \alpha _{2}\left( \tau
_{2p}^{\dagger }\Phi \right) \right] _{\alpha _{2}=0} &=&L_{np}\left[ \alpha
_{2}\left( \tau _{2p}^{\dagger }\Phi \right) \right] _{\alpha _{2}=0}, 
\end{eqnarray}
where $L_{nn}$ and $L_{np}$ (see eq.~(\ref{L})) contain, respectively,
the singlet neutron-neutron ($a_{nn}$, $R_{nn}$, $P_{nn}$) and triplet
neutron-proton ($ a_{np}$, $R_{np}$, $P_{np}$ ) parameters
\begin{eqnarray}
L_{nn} &=&\frac{\rho }{\sqrt{\mu }}\left[ \frac{1}{a_{nn}}+\frac{1}{2}
R_{nn}\left( \frac{\sqrt{\mu }\nu }{\rho }\right)
^{2}+P_{nn}R_{nn}^{3}\left( \frac{\sqrt{\mu }\nu }{\rho }\right) ^{4}\right]
, \\
L_{np} &=&\frac{\rho }{\sqrt{\mu }}\left[ \frac{1}{a_{np}}+\frac{1}{2}
R_{np}\left( \frac{\sqrt{\mu }\nu }{\rho }\right)
^{2}+P_{np}R_{np}^{3}\left( \frac{\sqrt{\mu }\nu }{\rho }\right) ^{4}\right]
\;.
\end{eqnarray}
Using the explicit form (\ref{dphi}) leads to a system of linear equations
for the coefficients $A_{i}$ 
\begin{eqnarray}
&&\left[ A_{1}\varphi _{\nu }^{\prime }(0)+A_{2}\left( \sigma _{nn}^{\dagger
}\tau _{2p}\right) \frac{4\varphi _{\nu }(\pi/3)}{\sqrt{3}}-A_{2}\left(
\sigma _{nn}^{\dagger }\tau _{p3}\right) \frac{4\varphi_\nu(\pi/3)}{\sqrt{3}}
\right] \nonumber \\ &&=L_{nn}A_{1}\varphi _{\nu }(0), \\
&&\left[ A_{2}\varphi _{\nu }^{\prime }(0)+A_{1}\left( \tau _{2p}^{\dagger
}\sigma _{nn}\right) \frac{4\varphi _{\nu }(\pi /3)}{\sqrt{3}}-A_{2}\left(
\tau _{2p}^{\dagger }\tau _{p3}\right) \frac{4}{\sqrt{3}}\varphi _{\nu }(\pi
/3)\right] \nonumber \\&&=L_{np}A_{2}\varphi _{\nu }(0) \;.
\end{eqnarray}
Again this system of equations has nontrivial solutions only when the
determinant of the corresponding matrix $M$ is zero, which gives the
eigenvalue equation
\begin{eqnarray}
&\det M =0 \;,&  \label{ddetm} \\
&M = &\nonumber \\
&\left( 
\begin{array}{c|c}
\varphi _{\nu }^{\prime }(0)-L_{nn}\varphi _{\nu }(0) & \left( \sigma
_{nn}^{\dagger }\tau _{2p}\right) \frac{4}{\sqrt{3}}\varphi _{\nu }(\frac{
\pi }{3})-\left( \sigma _{nn}^{\dagger }\tau _{p3}\right) \frac{4}{\sqrt{3}}
\varphi _{\nu }(\frac{\pi }{3}) \\ \hline
\left( \tau _{2p}^{\dagger }\sigma _{nn}\right) \frac{4}{\sqrt{3}}\varphi
_{\nu }(\frac{\pi }{3}) & \varphi _{\nu }^{\prime }(0)-\left( \tau
_{2p}^{\dagger }\tau _{p3}\right) \frac{4}{\sqrt{3}}\varphi _{\nu }(\frac{
\pi }{3})-L_{np}\varphi _{\nu }(0)
\end{array}
\right)& \nonumber
\;,\end{eqnarray}
where the cross-products of the spin functions can be easily calculated
from (\ref{ksit})
\begin{equation}
\sigma _{nn}^{\dagger }\tau _{2p}=-\frac{\sqrt{3}}{2},\;\sigma
_{nn}^{\dagger }\tau _{p3}=\frac{\sqrt{3}}{2},\;\tau _{2p}^{\dagger }\tau
_{p3}=-\frac{1}{2}\;.
\end{equation}
Without the regularizing terms, that is when effective ranges and shape
parameters are zero, the eigenvalue equation (\ref{ddetm}) has in the
limit $\rho \ll a$ an \emph{imaginary} root $\nu \cong 0.787i$ which
results in the Thomas collapse and the Efimov effect\cite{fed1pra}.

With non-zero effective ranges and shape parameters the equation
(\ref{ddetm}) has a \emph{real} root $\nu(\rho=0)$=0 and the Thomas
collapse is therefore removed. The three-body system then has a well
defined ground state. The Efimov effect can still exist in the limit
$a\rightarrow\infty$.

The large distance asymptotics is determined by the deuteron binding
energy and therefore is the same as for the quartet case, eq. (\ref{br2}).

Note that our regularization approach, contrary to \cite{adhikari,amorim}
and \cite {bedaque}, utilizes only two-body parameters and does not
require an additional three-body parameter.  In principle, we could also
introduce a three-body contact interaction in the hyper-radial equation
and use this additional parameter to further fine tune the model. However,
this is not the purpose of this paper.

\subsubsection{Adiabatic eigenvalues and results}

Contrary to the quartet eigenvalue, which only depends on the $np$
parameters, the doublet eigenvalue depends upon both $V_{np}$ and $V_{nn}$.
Fig. 4 shows the solutions of the eigenvalue equation (\ref{ddetm}) for all
possible combinations of $V_{np}$ and $V_{nn}$ from Table 1. The doublet
effective potential $(\lambda +4)/\rho ^{2}$ has an attractive pocket
where variation is somewhat larger than in the quartet case. One also
notices that all curves have basically the same shape, the only difference
being in the depth of the attractive pocket. This property is essential
for the discussion of the Phillips line in the next subsection.

\begin{figure}
\begin{center}
\input{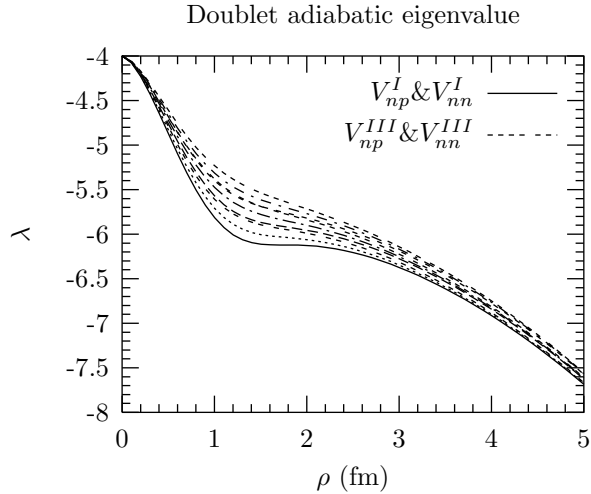}
\end{center}
\caption{
Doublet adiabatic eigenvalue $\lambda $ as function of $\rho
$ for different combinations of contact interactions from Table 1.  }
\end{figure}

Numerical solution of the hyper-radial equation (\ref{radial2}) with these
eigenvalues gives in each case one bound state, the triton, the energy of
which is shown in Table 3 together with the corresponding doublet $nd$
scattering length. Only the results for the corresponding potentials are
shown in the table. Other combinations give the results which fall somewhere
in between the shown results.

\begin{table}
\caption{
The triton ground state energy, $E_{t}$, and the doublet $nd$
scattering length, $^{2}a_{nd}$, for the nucleon-nucleon interactions from
Table 1.
}
\begin{center}
\begin{tabular}{|c|c|c|c|}
\hline
$V_{nn}$ & $V_{np}$ & $E_{t}$, MeV & $^{2}a_{nd}$, fm \\ 
\hline
I & I & -11.4 & 2.82 \\ 
II & II & -7.25 & -0.61 \\ 
III & III & -4.98 & -3.15 \\ 
\hline
\multicolumn{2}{|c|}{Exp.} & -8.48 & -0.65 \\
\hline
\end{tabular}
\end{center}
\end{table}

One can conclude that the contact interactions adequately describe the
triton and the low-energy doublet $nd$ scattering. The potentials II
even provide a good quantitative description similar to finite-range
potential models.

\subsubsection{Phillips line}

The Phillips line \cite{phillips} is an established relationship between
the triton binding energy and the doublet neutron-deuteron scattering
length -- different models for the nucleon-nucleon interactions with
the same low-energy properties produce different but correlated values
for these two quantities.  Graphically this correlation is customarily
represented as a curve (Phillips line) in the plot of the quartet $nd$
scattering length $^{4}a_{nd}$ versus the triton binding energy $B_{t}$.

The Phillips line for our model is shown on Fig.5 where we have plotted the
results for different combinations of the $np$ and $nn$ interactions from
Table 1. Also shown are the results from finite-range models \cite{chen} and
from the renormalized effective field theory with three-body force \cite
{eftline}.

\begin{figure}
\begin{center}
\input{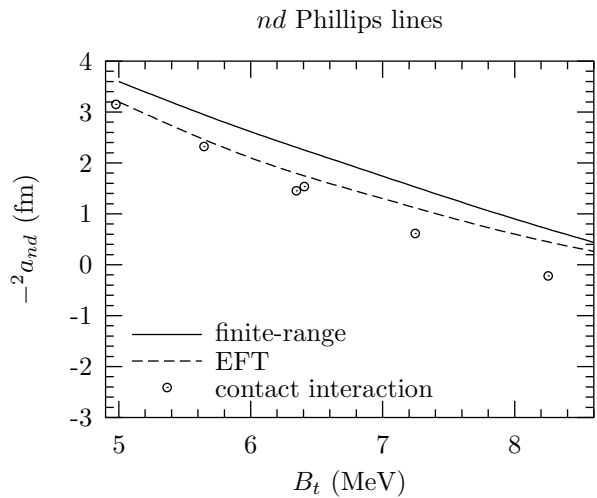}
\end{center}
\caption{
The Phillips plot of the doublet $nd$ scattering length $^2a_{nd}
$ versus the triton binding energy $B_{t}$. The points indicate the results
from different combinations of the contact interactions form Table 1. The
solid line represents the finite-range potential models \cite{chen}. The
dashed line represents the results from the renormalized effective field
theory with three-body force \cite{eftline}.
}
\end{figure}

Our Phillips line is reasonably close to but still somewhat different
from effective field theory curve. This indicates that the analytic
properties of these models are different.

Now the question to answer is why do the results fall on a one-parameter
curve while we in principle vary at least two independent ingredients --
the $nn$ and $np$ interactions? In our model the answer can be most easily
seen on Fig. 4 where shown are the eigenvalues for all combinations
of the $nn$ and $np$ interactions from Table~1. Due to the
analytic properties of the eigenvalue equation (\ref{ddetm}) all
curves start from $\lambda(\rho=0)=0$ and end up at $\lambda(\rho\gg
R)=-2mB_{d}\rho^2\hbar^2-4$. The shape of the curves is largely preserved
-- there are no crossings (with one exception) -- and the difference
is practically only in the depth of the attractive pocket. Therefore
we have got basically a \emph{one-parameter} family of the adiabatic
potentials, the parameter being, for example, the depth of the pocket. The
one-parameter family of adiabatic potentials would necessarily result
in a single curve in the Phillips plot where this parameter is changed
along the curve.

There are in fact two curves on Fig.~4 which, despite being almost
identical, do cross. Correspondingly these curves produce two close
points (around 6.3~MeV) on the Phillips plot, Fig.~5, which are slightly
off the line.

Note that in the effective field theory approach it is the three-body
cut-off parameter that is changed along the curve \cite{eftline}.
In our language that corresponds to fixing the two-body parameters,
thus fixing the adiabatic potential, and instead varying the strength
of the three-body force (one parameter).  This would, of course, also
lead to a single curve on the Phillips plot.

\section{Application to $\Lambda np$ system}

The $\Lambda np$ system has a weakly bound ground state, hypertriton, with
the spin 1/2 and the binding energy of 0.13$\pm$0.05~MeV relative to
the deuteron. Many investigations have been carried out with different
interaction models (see \cite{cobis,miyag} and references therein) some
of them including $\Lambda $-$\Sigma$ conversion \cite{miyag}. The
results exhibited considerable variations mostly due to uncertainties in
the lambda-nucleon interaction.

One might expect that a certain correlation exists between the $\Lambda
d$ scattering length and the hypertriton binding energy, similar to the
Phillips line for the $nnp$ system. The purpose of this chapter is to
establish the $\Lambda d$ Phillips line for our contact interaction
model and compare it with the finite-range model. We shall ignore
the $\Lambda$-$\Sigma$ conversion which is beyond the scope of
this paper. The mass of the $\Lambda$ particle is assumed to be
$m_\Lambda$=1116~MeV.

\subsection{Contact interactions}

For the $np$ subsystem we shall use the interaction II from Table 1. Due
to lack of experimental scattering data we cannot use the same procedure
for the $\Lambda N$ subsystem. Instead we simply fix the scattering
length and the effective range and vary the shape parameter around the
values similar to those of the $nn$ interaction. In \cite{cobis} a set of
low-energy parameters was found which gives a reasonable description of
the hypertriton within the finite-range three-body model.  We assume
that $p\Lambda $ and $n\Lambda $ interactions are identical and adopt
the scattering lengths and effective ranges in Table~\ref{hN}. They are
close to the set E1r in \cite{cobis}. The singlet $P_{s}$ and triplet
$P_{t}$ parameters are for simplicity assumed to be equal and are varied
in the range 0.08-0.11.

\begin{table}
\caption{Parameters of the $^{1}S_{0}$ and $^{3}S_{1}$ $\Lambda$-nucleon
contact interactions. The singlet and triplet shape parameters are
assumed to be equal and are varied in the range 0.08-0.11.}

\begin{center}
\begin{tabular}{|c|c|c|c|}
\hline
        & $a$, fm & $R$, fm & P \\
\hline
$^1S_0$ & 1.9     & 3.5     & 0.08-0.11 \\
\hline
$^3S_1$ & 1.5     & 3.7     & 0.08-0.11 \\
\hline
\end{tabular}
\end{center}\label{hN}
\end{table}

\subsection{Adiabatic eigenfunction and eigenvalue equation}

We assume that the $np$ pair is in the triplet state while $p\Lambda $ and $
n\Lambda $ pairs can be both in triplet and singlet states. We shall
therefore use the following spin-functions 
\begin{eqnarray}
\tau _{np} &=&\left( (n\otimes p)_{1}\otimes \Lambda \right) _{\frac{1}{2}}=
\sqrt{\frac{2}{3}}\left( \uparrow _{n}\uparrow _{p}\right) \downarrow
_{\Lambda }-\sqrt{\frac{1}{6}}\left( \uparrow
_{n}\downarrow _{p}+\downarrow _{n}\uparrow _{p}\right) \uparrow _{\Lambda },
\\
\sigma _{p\Lambda } &=&\left( (p\otimes \Lambda )_{0}\otimes n\right) _{
\frac{1}{2}}=\sqrt{\frac{1}{2}}\left( \uparrow _{p}\downarrow _{\Lambda
}-\downarrow _{p}\uparrow _{\Lambda }\right) \uparrow _{n}, \\
\tau _{p\Lambda } &=&\left( (p\otimes \Lambda )_{1}\otimes n\right) _{\frac{1
}{2}}=\sqrt{\frac{2}{3}}\left( \uparrow _{p}\uparrow _{\Lambda }\right)
\downarrow _{n}-\sqrt{\frac{1}{6}}\left( \uparrow _{p}\downarrow _{\Lambda
}+\downarrow _{p}\uparrow _{\Lambda }\right) \uparrow _{n}, \\
\sigma _{n\Lambda } &=&\left( (\Lambda \otimes n)_{0}\otimes p\right) _{
\frac{1}{2}}=\sqrt{\frac{1}{2}}\left( \uparrow _{\Lambda }\downarrow
_{n}-\downarrow _{\Lambda }\uparrow _{n}\right) \uparrow _{p}, \\
\tau _{n\Lambda } &=&\left( (\Lambda \otimes n)_{1}\otimes p\right) _{\frac{1
}{2}}=\sqrt{\frac{2}{3}}\left( \uparrow _{\Lambda }\uparrow _{n}\right)
\downarrow _{p}-\sqrt{\frac{1}{6}}\left( \uparrow _{\Lambda }\downarrow
_{n}+\downarrow _{\Lambda }\uparrow _{n}\right) \uparrow _{p}.
\end{eqnarray}
We assume that the wave-function is symmetric under exchange of the two
nucleons. Since under exchange $n\leftrightarrow p$ the spin functions
transform as 
\begin{equation}
\sigma _{p\Lambda }\leftrightarrow -\sigma _{n\Lambda },\tau _{p\Lambda
}\leftrightarrow \tau _{n\Lambda },\tau _{np}\leftrightarrow \tau _{np},
\end{equation}
the symmetric hyper-angular function (\ref{Phspin}) can be written as 
\begin{equation}\label{wfht}
\Phi =A\frac{\varphi _{\nu }(\alpha _{1})}{\sin 2\alpha _{1}}\tau _{np}+
\frac{\varphi _{\nu }(\alpha _{2})}{\sin 2\alpha _{2}}\left( B\sigma
_{p\Lambda }+C\tau _{p\Lambda }\right) +\frac{\varphi _{\nu }(\alpha _{3})}{
\sin 2\alpha _{3}}\left( -B\sigma _{n\Lambda }+C\tau _{n\Lambda }\right) .
\end{equation}
For this system we have three spin-projected boundary conditions
(\ref{bc3}), the triplet $np$, and singlet and triplet $N\Lambda $
\begin{eqnarray}
\tau _{np}^{\dagger }\left( \frac{\partial }{\partial \alpha _{1}}2\alpha
_{1}\Phi \right) _{\alpha _{_{1}}=0} &=&L^{(t)}_{np}\tau _{np}^{\dagger
}\left( 2\alpha _{1}\Phi \right) _{\alpha _{_{1}}=0}, \\
\sigma _{p\Lambda }^{\dagger }\left( \frac{\partial }{\partial \alpha _{2}}
2\alpha _{2}\Phi \right) _{\alpha _{_{2}}=0} &=&L_{\Lambda N}^{(s)}\sigma
_{p\Lambda }^{\dagger }\left( 2\alpha _{2}\Phi \right) _{\alpha _{_{2}}=0},
\\
\tau _{p\Lambda }^{\dagger }\left( \frac{\partial }{\partial \alpha _{2}}
2\alpha _{2}\Phi \right) _{\alpha _{2}=0} &=&L_{\Lambda N}^{(t)}\tau
_{p\Lambda }^{\dagger }\left( 2\alpha _{2}\Phi \right) _{\alpha _{_{2}}=0},
\end{eqnarray}
where $s$ and $t$ refer to singlet and triplet states and where, as
usual,
\begin{equation}
L_{ab}^{(c)}\equiv \frac{\rho }{\sqrt{\mu _{ab}}}\left( \frac{1}{a_{ab}^{(c)}
}+\frac{1}{2}R_{ab}^{(c)}\left( \frac{\nu \sqrt{\mu _{ab}}}{\rho }\right)
^{2}+P_{ab}^{(c)}\left( R_{ab}^{(c)}\right) ^{3}\left( \frac{\nu \sqrt{\mu
_{ab}}}{\rho }\right) ^{4}\right) \;,
\end{equation}
where $\mu_{ab}=(1/m)m_am_b/(m_a+m_b)$ is the reduced mass of particles
$a$ and $b$ in units of $m$.
The resulting system of linear equations for the coefficients $A$, $B$,
and $C$ is 
\begin{eqnarray}
A\varphi _{\nu }^{\prime }(0)+2\frac{2\varphi _{\nu }(\phi _{21})}{\sin
2\phi _{21}}\left( B\tau _{np}^{\dagger }\sigma _{p\Lambda }+C\tau
_{np}^{\dagger }\tau _{p\Lambda }\right)
=AL_{np}^{(t)}\varphi _{\nu }(0), \\
B\varphi _{\nu }^{\prime }(0)+A\frac{2\varphi _{\nu }(\phi _{21})}{\sin
2\phi _{21}}\sigma _{p\Lambda }^{\dagger }\tau _{np}+\frac{2\varphi _{\nu
}(\phi _{23})}{\sin 2\phi _{23}}\left( -B\sigma _{p\Lambda }^{\dagger
}\sigma _{n\Lambda }+C\sigma _{p\Lambda }^{\dagger }\tau _{n\Lambda }\right) 
\nonumber \\
=BL_{\Lambda N}^{(s)}\varphi _{\nu }(0), \\
C\varphi _{\nu }^{\prime }(0)+A\frac{2\varphi _{\nu }(\phi _{21})}{\sin
2\phi _{21}}\tau _{p\Lambda }^{\dagger }\tau _{np}+\frac{2\varphi _{\nu
}(\phi _{23})}{\sin 2\phi _{23}}\left( -B\tau _{p\Lambda }^{\dagger }\sigma
_{n\Lambda }+C\tau _{p\Lambda }^{\dagger }\tau _{n\Lambda }\right)
\nonumber  \\
=CL_{\Lambda N}^{(t)}\varphi _{\nu }(0).
\end{eqnarray}

And, finally, the eigenvalue equation 
\begin{eqnarray} 
&\det M =0 \;,& \label{heigen}\\
&M =& \nonumber \\
&\left( 
\begin{array}{c|c|c}
\varphi _{\nu }^{\prime }(0)-\varphi _{\nu }(0)L_{NN}^{(t)} & 2\frac{
2\varphi _{\nu }(\phi _{21})}{\sin 2\phi _{21}}\tau _{np}^{\dagger }\sigma
_{p\Lambda } & 2\frac{2\varphi _{\nu }(\phi _{21})}{\sin 2\phi _{21}}\tau
_{np}^{\dagger }\tau _{p\Lambda } \\ \hline
\frac{2\varphi _{\nu }(\phi _{21})}{\sin 2\phi _{21}}\sigma _{p\Lambda
}^{\dagger }\tau _{np} & 
\begin{array}{c}
\varphi _{\nu }^{\prime }(0)-\varphi _{\nu }(0)L_{\Lambda N}^{(s)} \\ 
-\frac{2\varphi _{\nu }(\phi _{23})}{\sin 2\phi _{23}}\sigma _{p\Lambda
}^{\dagger }\sigma _{n\Lambda }
\end{array}
& \frac{2\varphi _{\nu }(\phi _{23})}{\sin 2\phi _{23}}\sigma _{p\Lambda
}^{\dagger }\tau _{n\Lambda } \\ \hline
\frac{2\varphi _{\nu }(\phi _{21})}{\sin 2\phi _{21}}\tau _{p\Lambda
}^{\dagger }\tau _{np} & -\frac{2\varphi _{\nu }(\phi _{23})}{\sin 2\phi
_{23}}\tau _{p\Lambda }^{\dagger }\sigma _{n\Lambda } & 
\begin{array}{c}
\varphi _{\nu }^{\prime }(0)-\varphi _{\nu }(0)L_{\Lambda N}^{(t)} \\ 
+\frac{2\varphi _{\nu }(\phi _{23})}{\sin 2\phi _{23}}\tau _{p\Lambda
}^{\dagger }\tau _{n\Lambda }
\end{array}
\end{array}
\right),&   \nonumber
\end{eqnarray}
where the cross products of the spin-functions are 
\begin{eqnarray}
\tau _{np}^{\dagger }\sigma _{p\Lambda } &=&\frac{\sqrt{3}}{2},\tau
_{np}^{\dagger }\tau _{p\Lambda }=-\frac{1}{2},\tau _{np}^{\dagger }\sigma
_{n\Lambda }=-\frac{\sqrt{3}}{2}, \\
\sigma _{p\Lambda }^{\dagger }\sigma _{n\Lambda } &=&-\frac{1}{2},\sigma
_{p\Lambda }^{\dagger }\tau _{n\Lambda }=-\frac{\sqrt{3}}{2}, \\
\tau _{p\Lambda }^{\dagger }\sigma _{n\Lambda } &=&\frac{\sqrt{3}}{2},\tau
_{p\Lambda }^{\dagger }\tau _{n\Lambda }=-\frac{1}{2}.
\end{eqnarray}

\subsection{Eigenvalues and results}

\begin{figure}[tbp]
\begin{center}
\input{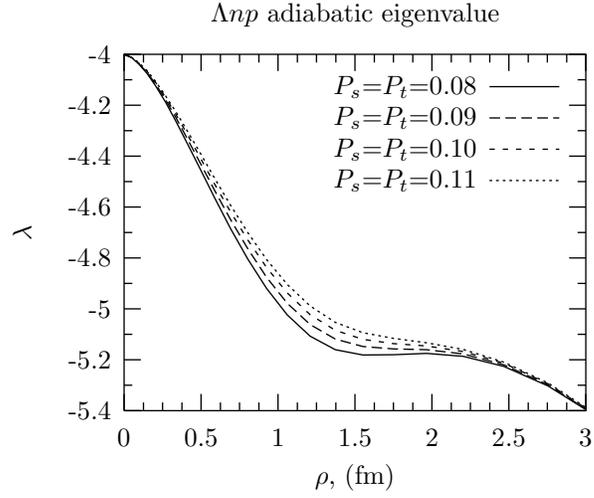}
\end{center}
\caption{Doublet adiabatic eigenvalue $\lambda$ as function of
$\rho$ for $\Lambda np$ system with contact interactions from
Table~\ref{hN}.}\label{llam}

\end{figure}

The solutions of the eigenvalue equation (\ref{heigen}) with parameters
from Table~\ref{hN} are shown on Fig.~\ref{llam}. Again the asymptotics
at both $\rho$=0 and $\rho\gg R$ are independent on the shape parameter.
The agreement between the curves in the asymptotic region is better
than in the triton case because here we keep the two-body scattering
lengths and effective ranges fixed (Table~\ref{hN}) while in the former
case they were slightly varied (Table~1).  The depth of the attractive
pocket, however, does show some variation with the shape parameter.

\begin{figure}[tbp]
\begin{center}
\input{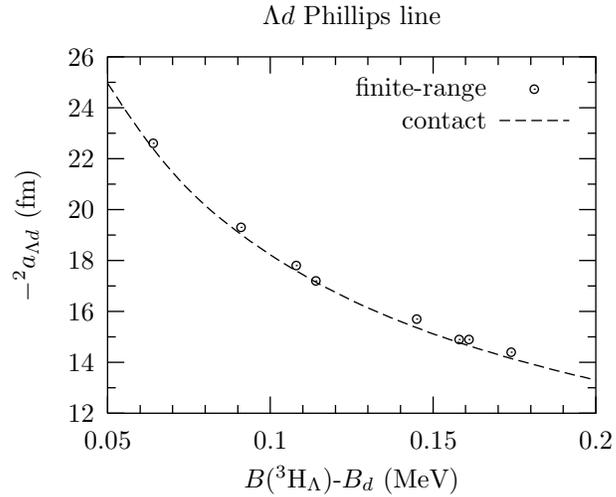}
\end{center}
\caption{The Phillips plot of the doublet $\Lambda d$ scattering
length $^2a_{\Lambda d}$ versus the triton binding energy relative
to deuteron ($B_d$=2.2~MeV). The curve is obtained from the contact
interaction model. The circles represent finite-range models
\protect\cite{cobis}. }\label{lpline}
\end{figure}

The corresponding hypertriton binding energies and doublet $\Lambda d$
scattering lengths are shown on the Phillips plot, Fig.~\ref{lpline},
together with the results from finite-range models \cite{cobis}.  The
agreement is even better than in the triton case, although the curve for
the contact interactions is still marginally below. Basically the contact
interaction model proves to be on a par with the finite-range model.

Note that this is not a simple repetition of the triton results as the
hypertriton has a very different structure -- the $\Lambda$ particle is
not identical to the neutron, which results in quite different angular
functions (see (\ref{wft}) and (\ref{wfht})), and also does not form a
bound state with the proton.

The better agreement reflects also the fundamental property of weakly
bound systems -- halos \cite{halos} -- of which the hypertriton is a
remarkable example, to be insensitive to the properties of the underlying
interaction models.

\section{Conclusion}

We have applied the recently formulated regularized zero-range model to
fermionic three-body systems $nnp$ and $\Lambda np$.  The model employs
low-energy two-body parameters, scattering length and effective range,
and the shape parameter.  The three-body problem is solved within the
hyper-spheric adiabatic approach where the zero-range model allows
analytic solutions for the adiabatic hyper-angular functions.

The large-distance asymptotics of the adiabatic potential is completely
determined by the scattering length and effective range and therefore
is model-independent. The shape parameter mostly determines the depth
of the attractive pocket and is thus important for the actual value of
the binding energy.

Our regularization approach does not require an additional
three-body parameter in contrast to the alternative momentum space
regularizations. The finite triton binding energy is obtained using only
two-body data, much like with the finite-range models.

Once the two-body parameters are fit to the nucleon-nucleon phase shifts
the model provides an adequate quantitative description of the $nnp$
system -- the bound state, triton, and the low-energy $nd$ scattering.
The established correlation between the doublet $nd$ scattering length and
the triton binding energy, the Phillips line, finds a natural explanation
within the model.

For the $\Lambda np$ system, with its weakly bound state, hypertriton,
the Phillips lines for zero-range and finite-range models are almost
identical, reflecting the fact that the topical weakly bound systems,
halos, are residing mostly at large distances, where the adiabatic
potential is independent upon the underlying potential model.  The
regularized zero-range model is therefore ideally suited for such systems.

The simple analytic functions of the model might serve as a convenient
Sturmian basis in other three-body applications.

In conclusion the regularized zero-range potential model can be
effectively used for quantitative descriptions of various three- and
many-body systems in discrete and continuum spectra.

\section{Acknowledgements}

We thank E.Nielsen and J.H.Macek for fruitful discussions.

\section{Hyper-spheric coordinates}

If $m_{i}$ and $\mathbf{r}_{i}$ refer to the $i$-th particle then the
hyper-radius $\rho $ and the hyper-angles $\alpha _{i}$ are defined in terms
of the Jacobi coordinates $\mathbf{x}_{i}$ and $\mathbf{y}_{i}$ as \cite{RR} 
\begin{eqnarray}
\mathbf{x}_{i} &=&\sqrt{\mu _{i}}(\mathbf{r}_{j}-\mathbf{r}_{k})\;,\;\mathbf{
y}_{i}=\sqrt{\mu _{(jk)i}}\left( \mathbf{r}_{i}-\frac{m_{j}\mathbf{r}
_{j}+m_{k}\mathbf{r}_{k}}{m_{j}+m_{k}}\right) \;,  \nonumber  \label{hyp} \\
\mu _{i} &=&\frac{1}{m}\frac{m_{j}m_{k}}{m_{j}+m_{k}}\;,\;\mu _{(jk)i}=\frac{1
}{m}\frac{m_{i}(m_{j}+m_{k})}{m_{i}+m_{j}+m_{k}} \\
\rho \sin (\alpha _{i}) &=&x_{i}\;,\;\;\rho \cos (\alpha _{i})=y_{i}\;, 
\nonumber
\end{eqnarray}
where $\{i,j,k\}$ is a cyclic permutation of \{1,2,3\} and $m$ is an
arbitrary mass. The set of angles $\Omega _{i}$ consists of the hyper-angle $
\alpha _{i}$ and the four angles $\mathbf{x}_{i}/|\mathbf{x}_{i}|$ and $
\mathbf{y}_{i}/|\mathbf{y}_{i}|$. The kinetic energy operator $T$ is then
given as 
\begin{eqnarray}
&&T=T_{\rho }+\frac{\hbar ^{2}}{2m\rho ^{2}}\Lambda \;,\;T_{\rho }=-\frac{
\hbar ^{2}}{2m}\left( \rho ^{-5/2}\frac{\partial ^{2}}{\partial \rho ^{2}}
\rho ^{5/2}-\frac{1}{\rho ^{2}}\frac{15}{4}\right) \;,  \label{def} \\
&&\Lambda =-\frac{1}{\sin (2\alpha _{i})}\frac{\partial ^{2}}{\partial
\alpha _{i}^{2}}\sin (2\alpha _{i})-4+\frac{l_{x_{i}}^{2}}{\sin ^{2}(\alpha
_{i})}+\frac{l_{y_{i}}^{2}}{\cos ^{2}(\alpha _{i})}\;,  \nonumber
\end{eqnarray}
where $\mathbf{l}_{x_{i}}$ and $\mathbf{l}_{y_{i}}$ are the angular momentum
operators related to $\mathbf{x}_{i}$ and $\mathbf{y}_{i}$.

\end{document}